\documentclass[aps,prb,reprint,superscriptaddress,floatfix]{revtex4-2}

\usepackage{multirow}
\usepackage{amsmath}
\usepackage{amssymb}
\usepackage{graphicx}
\usepackage{color}
\usepackage{array}
\usepackage{xspace}
\usepackage{bm}
\usepackage{xcolor}

\begin{document}
\title{Multi-band $s_{++}$ superconductivity in $\textrm{V}_{3}\textrm{Si}$
determined \\
 from the response to a controlled disorder }
\author{Kyuil Cho}
\affiliation{Ames Laboratory, Ames, IA 50011, USA}
\affiliation{Department of Physics \& Astronomy, Iowa State University, Ames, IA
50011, USA}
\author{M. Ko\'{n}czykowski}
\affiliation{Laboratoire des Solides Irradi\'{e}s, CNRS UMR 7642 \& CEA-DSM-IRAMIS,
\'{E}cole Polytechnique, F-91128 Palaiseau cedex, France}
\author{S. Ghimire}
\affiliation{Ames Laboratory, Ames, IA 50011, USA}
\affiliation{Department of Physics \& Astronomy, Iowa State University, Ames, IA
50011, USA}
\author{M. A. Tanatar}
\affiliation{Ames Laboratory, Ames, IA 50011, USA}
\affiliation{Department of Physics \& Astronomy, Iowa State University, Ames, IA
50011, USA}
\author{Lin-Lin Wang}
\affiliation{Ames Laboratory, Ames, IA 50011, USA}
\author{V. G. Kogan}
\affiliation{Ames Laboratory, Ames, IA 50011, USA}
\author{R. Prozorov}
\email{prozorov@ameslab.gov}

\affiliation{Ames Laboratory, Ames, IA 50011, USA}
\affiliation{Department of Physics \& Astronomy, Iowa State University, Ames, IA
50011, USA}

\date{December 7, 2021}

\begin{abstract}
The London penetration depth, $\lambda(T)$, was measured in a single
crystal V$_{3}$Si. The superfluid density obtained from this measurement
shows a distinct signature of two almost decoupled superconducting
gaps. This alone is insufficient to distinguish between $s_{\pm}$
and $s_{++}$ pairing states, but it can be achieved by studying the
effect of a controlled non-magnetic disorder on the superconducting
transition temperature, $T_{c}$. For this purpose, the same $\text{V}_{3}\text{Si}$
crystal was sequentially irradiated by 2.5 MeV electrons three times,
repeating the measurement between the irradiation runs. A total dose
of 10 C/cm$^{2}$ ($6.24\times10^{19}$ electrons/$\textrm{cm}^{2}$)
was accumulated, for which $T_{c}$ has changed from 16.4 K in a pristine
state to 14.7 K (9.3 $\%$). This substantial suppression is impossible
for a single isotropic gap, but also it is not large enough for a sign-changing
$s_{\pm}$ pairing state. Our electronic band-structure calculations
show how five bands crossing the Fermi energy can be naturally grouped
to support two effective gaps, not dissimilar from the iron pnictides
physics. We analyze the results using two-gap models for both, $\lambda(T)$
and $T_{c}$, which describe the data very well. Thus, the experimental
results and theoretical analysis provide strong support for an $s_{++}$
superconductivity with two unequal gaps, $\Delta_{1}\left(0\right)\approx2.53\;\textrm{meV}$
and $\Delta_{2}\left(0\right)\approx1.42\;\textrm{meV}$, and a very
weak inter-band coupling in $\text{V}_{3}\text{Si}$ superconductor. 
\end{abstract}
\maketitle

\section{Introduction}

At the time of its discovery in 1953 \cite{Hardy1953}, a cubic (A15
structure) $\text{V}_{3}\text{Si}$ compound had the highest superconducting
transition temperature, around 17 K. Despite showing a clear exponential
attenuation of all thermodynamic quantities upon cooling towards $T=0$,
which signaled a fully gapped Fermi surface, most of spectroscopic
\cite{Blezius1986,Wu1994Hyperfine_V3Si_Lambda_0}, transport \cite{Testardi1977PRB,Viswanathan1978PRB_neutron_V3Si}
and thermodynamic measurements \cite{Wu1994Hyperfine_V3Si_Lambda_0,Khlopkin1999,Kogan2009PRB_gamma_model,Tanaka2010JPCS_V3Si_pressure}
showed unconventional behavior or at least some unusual features.
Such behavior can be associated with a peculiar electronic band-structure
showing Van Hove singularities in the density of states (DOS) close
to the Fermi level \cite{GOLDBERG1972,Klein1978,Wu1994Hyperfine_V3Si_Lambda_0,Bok2012}.
While this certainly plays an important role, now we know that a multi-gap
superconductivity is needed as well to understand the measurements.
Here we focus on a multi-band, multi-gap nature of superconductivity
in this fascinating material.

While $\textrm{MgB}_{2}$ \cite{Nagamatsu2001} is commonly accepted
as the first confirmed two-gap superconductor \cite{Bouquet2001,Kortus2001,Budko2001},
the multi-band superconductivity was studied much earlier, albeit
only theoretically. Soon after the development of the microscopic
model of superconductivity \cite{BCS1957} the possibility of ``overlapping
bands'' was studied \cite{Moskalenko1959,Suhl1959,Geilikman1967},
eventually leading to a general description of multi-band superconductivity
\cite{Mazin1993,Golubov1994,Golubov1995,Golubov1997}, in particular
the effects of disorder \cite{Golubov1997}. Nevertheless, before
$\textrm{MgB}_{2}$, there was no attempt to interpret the unusual
properties of $\text{V}_{3}\text{Si}$ through the prism of multi-band
effects. The observations of the unconventional London penetration
depth \cite{Blezius1986,Wu1994Hyperfine_V3Si_Lambda_0}, anisotropic
upper critical field, $H_{c2}$ \cite{Khlopkin1999}, an unexpectedly
large decrease of $T_{c}$ with nonmagnetic disorder, either after
neutron irradiation \cite{Viswanathan1978,Viswanathan1978PRB_neutron_V3Si}
or naturally present in real material \cite{Orlando1979}, a large
$T_{c}/T_{F}\sim0.01$ ratio ($T_{F}$ is Fermi temperature) \cite{Wu1994Hyperfine_V3Si_Lambda_0}
and a variety of vortex lattice configurations \cite{Kogan_1997}
all pointed to an unconventional behavior of a confirmed $s-$wave
superconductor. Surely, modern re-interpretation of many of these
results is consistent with, if not fully explained, by multi-gap superconductivity.
Therefore, retrospectively, $\textrm{V}_{3}\textrm{Si}$ is a much
earlier than $\textrm{MgB}_{2}$, ``the first'' two-band superconductor.

Experimental observation of a two-gap superconducting state relies on a substantial decoupling between the two bands and a substantial
difference between them either in terms of dimensionality, electronic
properties, pairing mechanism and/or scattering rates \cite{Kogan2009PRB_gamma_model,Prozorov2011RPP_review,Zehetmayer2014SST_single_band_V3Si,MgB2Symmetry2019}.
In terms of more recent measurements, when multi-band superconductivity
became widely accepted and discussed, circa 2001, while some reports
support single-gap conventional $s-$wave BCS superconductivity in
$\textrm{V}_{3}\textrm{Si}$ \cite{Zehetmayer2014SST_single_band_V3Si},
many more experimental and theoretical studies point to two distinct
energy gaps in this material \cite{Nefyodov2005EPL_two-gap_V3Si,Kogan2009PRB_gamma_model,Perucchi2010PRB_V3Si_two_gap_optical,Tanaka2012JPSJ_V3Si_pressure}.
There is a complication, though. Perhaps due to a variation of stoichiometry,
atomic disorder or extremely strain-sensitive structure of Van Hove
singularities in the vicinity of the Fermi level, $\textrm{V}_{3}\textrm{Si}$
samples show a spread of behaviors, especially in the properties related
to a two-gap superconductivity \cite{Testardi1967,Testardi1977PRB,Klein1978,Viswanathan1978,Nefyodov2005EPL_two-gap_V3Si,Tanaka2010JPCS_V3Si_pressure,Bok2012}.
Furthermore, establishing a multi-band nature from thermodynamic measurements
is necessary but insufficient for the microscopic understanding of
superconductivity, because the order parameter enters thermodynamic
quantities in the even powers, therefore the gaps of the same or opposite
signs on different bands contribute similarly \cite{Kogan2009PRB_gamma_model,Prozorov2011RPP_review,Carrington2011,Kogan2016}.
In this situation, phase-sensitive experiments are needed, but it
is often difficult to implement experimentally \cite{vanHarlingenRMP1995,Golubov1995,Golubov2013}.
While in high$-T_{c}$ cuprates direct measurements that depend on
the phase variation along the Fermi surface have provided a definitive
proof of a sign-changing $d-$wave order parameter \cite{vanHarlingenRMP1995},
in multi-band iron-based superconductors, a similar simple arrangement
in real space is not possible, and more complicated approaches are
needed \cite{Golubov2013}. The interpretations of more complex phase-sensitive
experiments, such as quasiparticles interference, are not straightforward
either \cite{Sykora2011,Hanaguri2010}.

Scattering off non-magnetic impurities is a phase-sensitive method,
albeit indirect, that was successfully used in iron-based superconductors
to probe the sign-changing multi-band $s_{\pm}$ order parameter \cite{Efremov2011,FeRh122_PRL2018,Cho2018SST_review_e-irr}.
As we discuss in Section III.E, the suppression of $T_{c}$ formally
depends on the Fermi surface averaging of the order parameter in the
first power, $\left\langle \Delta\left(\mathbf{k}\right)\right\rangle _{FS}$,
which is sign-sensitive. For example, isotropic $s-$wave angular part
averages to 1, but $d-$wave averages to 0. Of course, more than one
measurement is needed for different levels of scattering in the system.
In our approach, simultaneous measurements of normal state resistivity
(to characterize the amount of introduced disorder), the superconducting
transition temperature, $T_{c}$ (phase-sensitive measurement), and
low-temperature variation of the London penetration depth, $\lambda\left(T\right)$
(to estimate the anisotropy of the order parameter amplitude), provide
enough information to make that conclusion. Here we show that this scheme can be applied to prove the existence of two distinct gaps
of the same sign, or $s_{++}$ order parameter, in the title material,
$\textrm{V}_{3}\textrm{Si}$. The utility of such an approach was extended
significantly by recent theoretical analysis of the impurity scattering
in superconductors with non-trivial multi-band structure \cite{Teknowijoyo2018,Krenkel2021}.
For example, it is possible to have a singlet unconventional pairing
with a sign-changing superconducting order parameter, yet fully gapped
Fermi surface, similar to what we uncover here. However, the electronic
band-structure should support such an unconventional scenario in a first
place. We note that nodeless unconventional superconductivity has
been studied in a context of triplet pairing, such as $p-$wave, which
shows a variety of nodal and nodeless behaviors depending on the material
and experimental conditions, for example, in some heavy-fermion superconductors
\cite{Gross1986,Gro_Alltag1991}. However, most superconductors have
singlet pairing states and the studies of the effects of a controlled
disorder is a powerful tool to study unconventional and exotic states,
including multi-gap superconductivity \cite{Cho2018SST_review_e-irr}.
For example, a similar combination of London penetration depth measurements
of electron-irradiated samples was used to study disorder-driven transitions
of the superconducting gap \cite{Mizukami2014}, the interplay of ferromagnetism
and superconductivity \cite{Ghimire2021}, proving fully-gapped superconductivity
in a heavy-fermion superconductor \cite{Takenaka2017}, and following
the doping evolution of the order parameter \cite{SciAdvBaK122lambda2016}.
If we include other types of irradiation, many studies employed neutrons
and protons to induce non-magnetic disorder. Such disorder was used
to induce a two-gap to a single-gap crossover in MgB$_{2}$ \cite{Putti2006},
trace the evolution of $s_{\pm}$ symmetry in iron-pnictides \cite{Schilling2016}
and studying its cross-over to $s_{++}$ state \cite{FeRh122_PRL2018},
or significantly suppress the superfluid density \cite{Kim2012}.
Evidently, a controlled disorder in conjunction with thermodynamic measurements
is a well-established approach to tune and probe the superconducting
state.

\section{Experimental}

Our $\textrm{V}_{3}\textrm{Si}$ crystals with $T_{c}\approx16.4$
K were cut out of a ``master boule'' single crystal studied previously,
for example in Refs.\cite{Yethiraj1999,Yethiraj2005} and references
therein. The resistivity above $T_{c}$ of pristine samples was in
range of $5-10\:\mathrm{\mu\Omega}\cdot\text{cm}$, consistent with
the previous reports \cite{Orlando1979,Yethiraj2005,Yethiraj1999}.
The samples were of sub-mm size. In particular, the crystal used in
electron irradiation study was $0.73\times0.62\times0.2\;\textrm{mm}^{3}$.

The variation of the in-plane London penetration depth, $\Delta\lambda(T)$,
was measured using a self-oscillating tunnel-diode resonator (TDR)
technique \cite{Carrington2011,Prozorov2006SST,Prozorov2011RPP_review}.
The TDR circuit resonates approximately at 14 MHz, and the frequency
shift is measured with a precision better than one part per billion
(ppb). Its inductor coil generates ac magnetic field, $H_{ac}<20~\text{mOe}$,
so that the sample is always in the Meissner state at temperatures
of interest. Details of the technique and its principles are given
in Ref.\cite{VanDegrift1975RSI,Prozorov2000PRB,Prozorov2000a} and
the detailed calibration procedure is described in Refs.\cite{Prozorov2021,Prozorov2000PRB}.
The sample was mounted on a 1 mm diameter sapphire rod and inserted
into a 2 mm diameter inductor coil. The coil and the sample were mounted
in a vacuum inside a $^{3}\textrm{He}$ cryostat. The TDR circuit was
actively stabilized at 5 K, and the sample was controlled from 0.4
K and up by independent LakeShore controllers. It is straightforward
to show that the change of the resonant frequency when a sample is
inserted into the coil is proportional to the sample magnetic susceptibility
as long as the change of the total inductance is small and one can
expand, $\Delta f/f_{0}\approx\varDelta L/2L_{0}$ where $2\pi f_{0}=1/\sqrt{CL_{0}}$
with sub-index $"0"$ referring to an empty resonator. The coefficient
of proportionality that includes the demagnetization correction is
measured directly by pulling the sample out of the resonator at the
base temperature \cite{Prozorov2021}.

\begin{figure}[tb]
\includegraphics[width=0.9\linewidth]{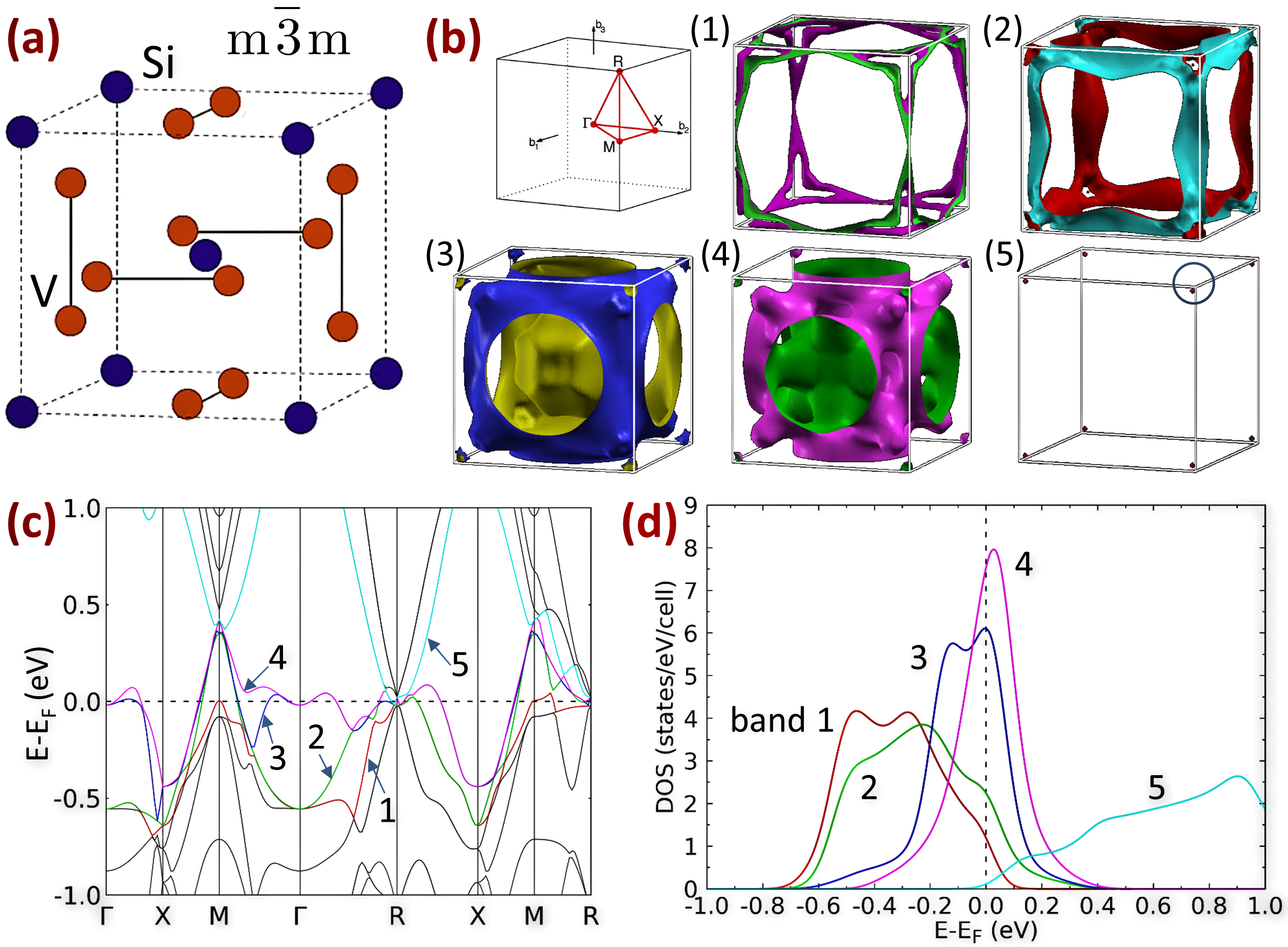}
\caption{\textbf{(a)} $\text{V}_{3}\text{Si}$ unit cell with two formula units,
$Z=2$. \textbf{(b)} Brillouin zone (BZ) and Fermi surfaces (FS) of
the five different bands crossing the Fermi level ($E_{F}$) for $\text{V}_{3}\text{Si}$.
The band numbers correspond to the numbers in the text. (The small
FS of Band 5 is circled for clarity.) \textbf{(c)} Energy band dispersion
along the high symmetry directions of BZ and \textbf{(d)} Partial
density of states as function of energy for the five bands crossing
$E_{F}$.}
\label{fig1:BS} 
\end{figure}

The low-temperature 2.5 MeV electron irradiation was performed at
the SIRIUS Pelletron facility of the Laboratoire des Solides Irradi\'{e}s
(LSI) at the \'{E}cole Polytechnique in Palaiseau, France. The acquired
irradiation dose is conveniently expressed in C/cm$^{2}$ and measured
directly as a total charge accumulated behind the sample by a Faraday
cage. Therefore, 1 C/cm$^{2}\approx6.24\times10^{18}$ electrons/cm$^{2}$.
In the experiment, the London penetration depth was measured, then
the sample was irradiated, and the cycle repeated. The irradiation
was carried out with the sample immersed in liquid hydrogen at about
20 K. Low-temperature irradiation is needed to slow down recombination
and migration of defects. Upon warming up to room temperature, a quasi-equilibrium population of atomic vacancies remains due to a substantial difference
in the migration barriers between vacancies and interstitials. An
example of such incremental irradiation/measurement sequence showing
the resistivity change measured in-situ, as well as the annealing
after warming up, is given elsewhere \cite{Prozorov2014PRX_BaRu}.
In the present case, the sample was dispatched between the lab and
the irradiation facility for the measurements and irradiation, and
then the sequence was repeated until the sample had accumulated a
substantial dose of 10 C/cm$^{2}$ $\approx6.24\times10^{19}$ electrons/cm$^{2}$.
Further information on the physics of electron irradiation can be found elsewhere \cite{Damask1963,THOMPSON1969}.

Density functional theory (DFT) with Perdew-Burke-Ernzerhof (PBE)
exchange-correlation functional \cite{Perdew1996} has been used to
calculate the band structure of $\textrm{V}_{3}\textrm{Si}$ at the
experimental lattice constant of $a=4.741\:\text{Å}\;$ \cite{Tanaka2011}.
The DFT calculations have been done in VASP \cite{Kresse1996} using
projected augmented wave method and a plane-wave basis set with a
kinetic energy cutoff of 246 eV. The charge density is converged on
a ($8\times8\times8$) Monkhorst-Pack $k-$point mesh, including the
$\Gamma-$point. For the Fermi surface (FS) calculations, a much denser
($30\times30\times30$) $k-$point mesh is used. The Fermi velocity
for each band has been calculated by the derivative of the DFT band
dispersion, i.e., group velocity, and then averaged over the Fermi
surface of each band in the Brillouin zone (BZ), the same method as
employed previously \cite{Belashchenko2001,Torsello2020}.

\section{Results and Discussion}

\subsection{Electronic band-structure}

\begin{table}[tbh]
\caption{
\label{table:BS}
Electronic band structure parameters relevant to
the $\gamma-$model fitting. The bands are naturally grouped in two
effective bands $I$ and $II$. The calculated parameter $\gamma=\left(n_{1}v_{1}^{2}+n_{2}v_{2}^{2}\right)/\sum_{i=1}^{5}n_{i}v_{i}^{2}=n_{I}v_{I}^{2}/\left(n_{I}v_{I}^{2}+n_{II}v_{II}^{2}\right)=0.109$,
to be compared with the experimental best fit, $\gamma=0.175$. The
effective quantities remapped on two effective bands are shown in
the last two columns.}
\begin{ruledtabular}
\begin{tabular}{|c|c|c|c|c|c|}
\multirow{2}{*}{band}  & $v_{F}^{2}\times10^{15}$  & DOS  & Two  & $v_{F}^{2}\times10^{15}$  & DOS\tabularnewline
\hline 
 & $\left(\textrm{cm}/\textrm{s}\right)^{2}$  & states/eV/cell  & bands  & $\left(\textrm{cm}/\textrm{s}\right)^{2}$  & st./eV/cell\tabularnewline
\hline 
\hline 
1  & $2.22$  & $1.22$  & \multirow{2}{*}{$I$} & \multirow{2}{*}{3.69} & \multirow{2}{*}{1.74}\tabularnewline
\cline{1-3} \cline{2-3} \cline{3-3} 
2  & $4.48$  & $2.26$  &  &  & \tabularnewline
\hline 
3  & $7.11$  & $6.11$  & \multirow{2}{*}{$II$} & \multirow{2}{*}{7.70} & \multirow{2}{*}{6.80}\tabularnewline
\cline{1-3} \cline{2-3} \cline{3-3} 
4  & $8.18$  & $7.48$  &  &  & \tabularnewline
\hline 
5  & $0.00315$  & $0.11$  & $\times$  & \multicolumn{1}{c}{} & \multicolumn{1}{c}{}\tabularnewline
\cline{1-4} \cline{2-4} \cline{3-4} \cline{4-4} 
\end{tabular}
\end{ruledtabular}
\end{table}

$\textrm{V}_{3}\textrm{Si}$ has a primitive cubic crystal structure
in space group 223 ($\text{Pm}\overline{3}\text{m}$) with V sitting
at 6c and Si at 2a positions as shown in Fig.\ref{fig1:BS}(a). The
band structure of $\textrm{V}_{3}\textrm{Si}$, Fig.\ref{fig1:BS}(c)
has flat pieces along $\Gamma$-X, $\Gamma$-M and $\Gamma$-R directions,
which is similar to $\text{Nb}_{3}\text{Sn}$, another A15 superconductor
with the same group of elements. There are five bands crossing the
$E_{F}$ as highlighted in different colors in Fig.\ref{fig1:BS}(c).
The corresponding partial densities of state (DOS) of these bands
are plotted in Fig.\ref{fig1:BS}(d) and summarized in Table~\ref{table:BS}.
Among them, bands 1 and 2 are hole bands with states gathering along
the M-R direction, the edges of the cubic Brillouin zone (BZ) (see
Fig.\ref{fig1:BS}(b)((1) and (2)). On the other hand, band 5 has
a very small electron pocket around the R point, and the contribution
to DOS is negligibly small. In contrast, bands 3 and 4 are dominant
in DOS at $E_{F}$, which corresponds to most of the flat band contributions
around the $\Gamma$ point as seen in Fig.\ref{fig1:BS}(c). The three-dimensional
(3D) Fermi surfaces (FS) in Fig.\ref{fig1:BS}(b)((3) and (4)) show
complex FS for both bands 3 and 4, which have multiple sheets at $E_{F}$.
Analysis of Fig.\ref{fig1:BS}(b) suggests that the five bands can
be naturally grouped in two effective ones. Specifically, bands 1
and 2 are well separated in energy from bands 3 and 4 at intermediate
$\mathbf{k}-$values inside the BZ, making the inter-band transitions
improbable. Furthermore, bands 1 and 2 are much closer in energy,
and this is also true for bands 3 and 4, but at different $\mathbf{k}$.
This suggests grouping bands 1 and 2 into an effective band $I$,
bands 3 and 4 into another effective band $II$, and discarding negligible-DOS
band 5. Electronic parameters of all five bands are reported in Table.\ref{table:BS}.
The ``effective'' parameters of two effective bands, $I$ and $II$,
are given in the last two columns. The multi-band average for band
$I$ is $v_{F,I}^{2}=(n_{1}v_{F,1}^{2}+n_{2}v_{F,2}^{2})/(n_{1}+n_{2})$,
and similar for the effective band $II$. As we explain in the two-band
$\gamma-$model Section III.C, the relative contribution of each band
to the superfluid density, $\rho_{s}=\gamma\rho_{I}+\left(1-\gamma\right)\rho_{II},$is
given by the parameter, $\gamma=n_{I}v_{I}^{2}/\left(n_{I}v_{I}^{2}+n_{II}v_{II}^{2}\right)$
(hence, $\gamma-$model). As shown in Table \ref{table:BS}, we estimate
$\gamma=0.109$, which is quite close to the experimental $\gamma=0.109$,
discussed in the next Section III.C.

The high DOS at $E_{F}$ in $\textrm{V}_{3}\textrm{Si}$ indicates
electronic instability, consistent with literature reports \cite{Bok2012,Klein1978,Wu1994Hyperfine_V3Si_Lambda_0}.
Although one way to reduce such instability is to promote an exchange
splitting, giving a magnetic solution at the DFT level, experimentally
$\textrm{V}_{3}\textrm{Si}$ is not magnetic. Another way to lift
the electronic instability is through the electron-phonon coupling.
Similar band structure with flat bands in $\text{Nb}_{3}\text{Sn}$
is susceptible to lattice distortion by a phonon
mode \cite{Sadigh1998}, indicating a strong electron-phonon coupling
in such compounds, hence obvious connection to superconductivity.
Thus, the non-magnetic electronic band structure of $\textrm{V}_{3}\textrm{Si}$
provides important microscopic details for superconductivity models,
such as DOS at $E_{F}$ and Fermi velocity, which have been used successfully
for MgB$_{2}$, the first proven two-band superconductor. In fact,
we calculated the electronic band structure of $\text{MgB}_{2}$ as a
benchmark to compare with with the original $\gamma-$model \cite{Kogan2009PRB_gamma_model,MgB2Symmetry2019}
and one of the first DFT calculations of a two-gap system \cite{Belashchenko2001},
and obtained similar results.

\subsection{London penetration depth and superfluid density}

\begin{figure}[tb]
\includegraphics[width=0.9\linewidth]{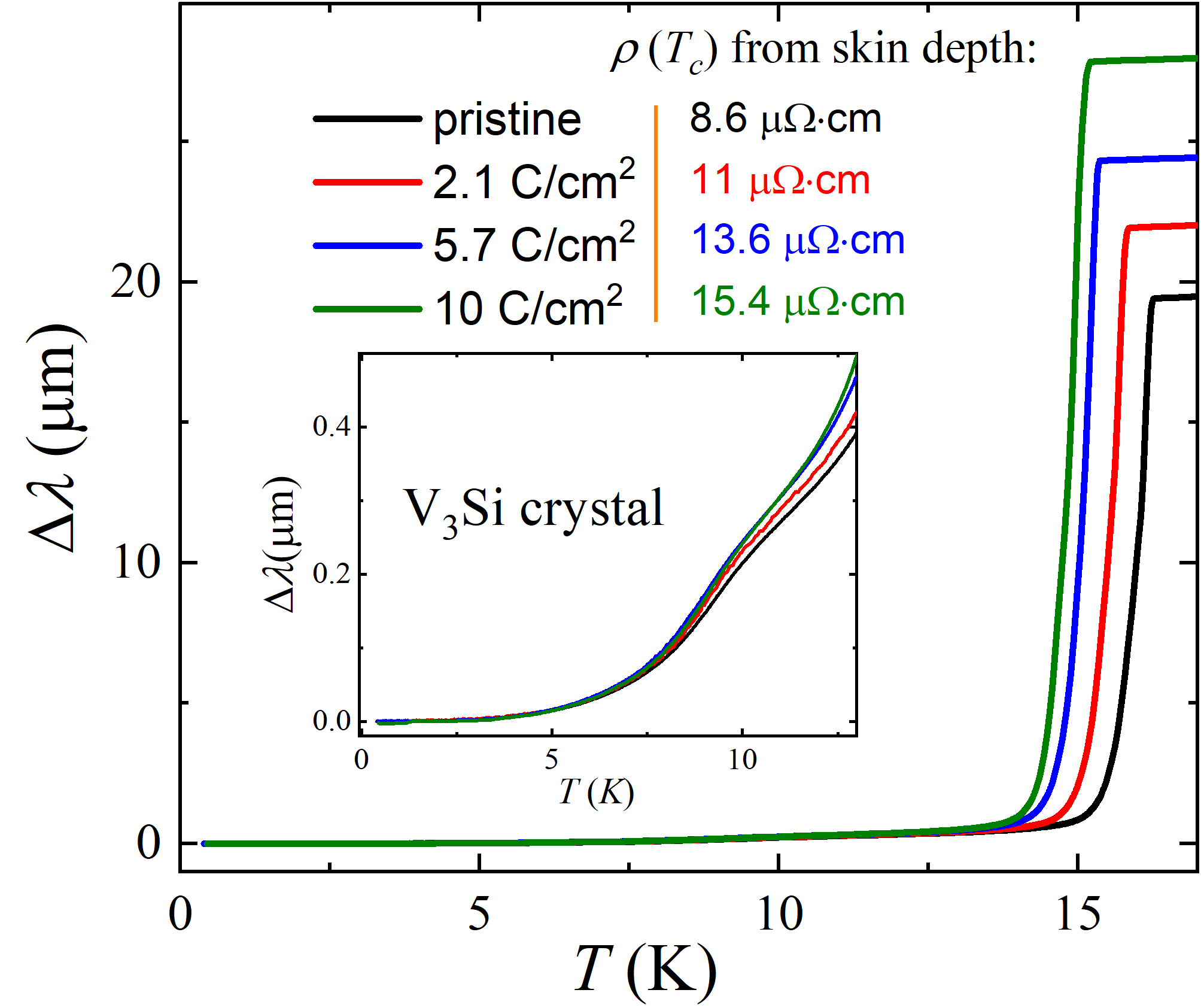}
\caption{Temperature dependent London penetration depth measured in a V$_{3}$Si
single crystal in pristine state and after three doses of electron
irradiation. The inset zooms at the low temperature region showing
a clear signature of a second gap developing at around 10 K.}
\label{fig2:lambda} 
\end{figure}

We begin by examining the superfluid density obtained from the measured
London penetration depth. Figure~\ref{fig2:lambda} shows temperature
dependent variation of London penetration depth, $\varDelta\lambda\equiv\lambda\left(T\right)-\lambda\left(T_{min}\right)$,
with the increasing dose of electron irradiation. Considering the
exponential low-temperature behavior, we can safely assume that $\lambda\left(T_{min}=0.4\:\textrm{K}\right)\approx\lambda\left(0\right)$
and then the normalized superfluid density is calculated as $\rho_{s}=\left(\varDelta\lambda\left(0\right)/\varDelta\lambda\left(T\right)\right)^{2}=\left(1+\varDelta\lambda/\lambda\left(0\right)\right)^{-2}$.
The inset in Fig.\ref{fig2:lambda} zooms at the low-temperature region.
There is a clear, almost knee-like feature in $\varDelta\lambda\left(T\right)$
around 10 K, which we now know is expected for a two-gap superconductor
with different and weakly-coupled gaps \cite{Prozorov2011RPP_review}.
Similar features were reported in high quality $\textrm{V}_{3}\textrm{Si}$
crystals before and not surprisingly was interpreted as a definitive
evidence of a two-gap superconductivity \cite{Kogan2009PRB_gamma_model,Nefyodov2005EPL_two-gap_V3Si,Kogan2009PRB_gamma_model,Nefyodov2005EPL_two-gap_V3Si}.
The temperature of this kink-like feature is suppressed upon irradiation
approximately at the same rate as $T_{c}$, signaling that both gaps
change at a similar rate. Furthermore, in Fig.\ref{fig2:lambda},
normal state values above $T_{c}$ are determined by the skin depth.
They increase upon irradiation due to the increase of residual resistivity
following the Matthiessen's rule \cite{Matthiessen1864a}.
Importantly, the superconducting transition temperature, $T_{c}$,
is monotonically and noticeably suppressed from 16.4 K to 14.7 K (9.3$\%$).
The upper cut-off at $T_{c}$ is determined by the normal-metal skin
depth, which allows us to estimate the resistivity in a contact-less
way using $\rho=\mu_{0}\pi f\delta^{2}$ where $\delta\left(T_{c}\leftarrow T\right)\approx2\lambda\left(T\rightarrow T_{c}\right)$
and $\lambda\left(T\right)=\varDelta\lambda\left(T\right)+\lambda\left(0\right)$,
where $\lambda(0)=130\;\textrm{nm}$ from Ref.\cite{Wu1994Hyperfine_V3Si_Lambda_0,Testardi1977PRB}.
The extracted resistivity values are 8.6, 11, 13.6, and 15.4 $\mu\Omega$cm
for 0, 2.1, 5.7 and 10 C/cm$^{2}$ electron irradiation doses, respectively.
These values appear to be quite comparable with the literature \cite{Testardi1977PRB,Viswanathan1978PRB_neutron_V3Si,Orlando1979}.
We note that in a large body of work on $\mathrm{V}_{3}\mathrm{Si}$,
a spread of $\lambda(0)$ values ranging from 83 nm to 230 nm can
be found. They were obtained using different methods, and in samples
of different forms (crystal vs. polycrystalline) and purity \cite{Hanaguri1995,Greytak1964,Muto1979}.
The value we use is within the statistical maximum of the current
literature values. Importantly, our results and conclusions are independent
of the particular value of $\lambda(0)$. 

\begin{figure}[tb]
\includegraphics[width=0.9\linewidth]{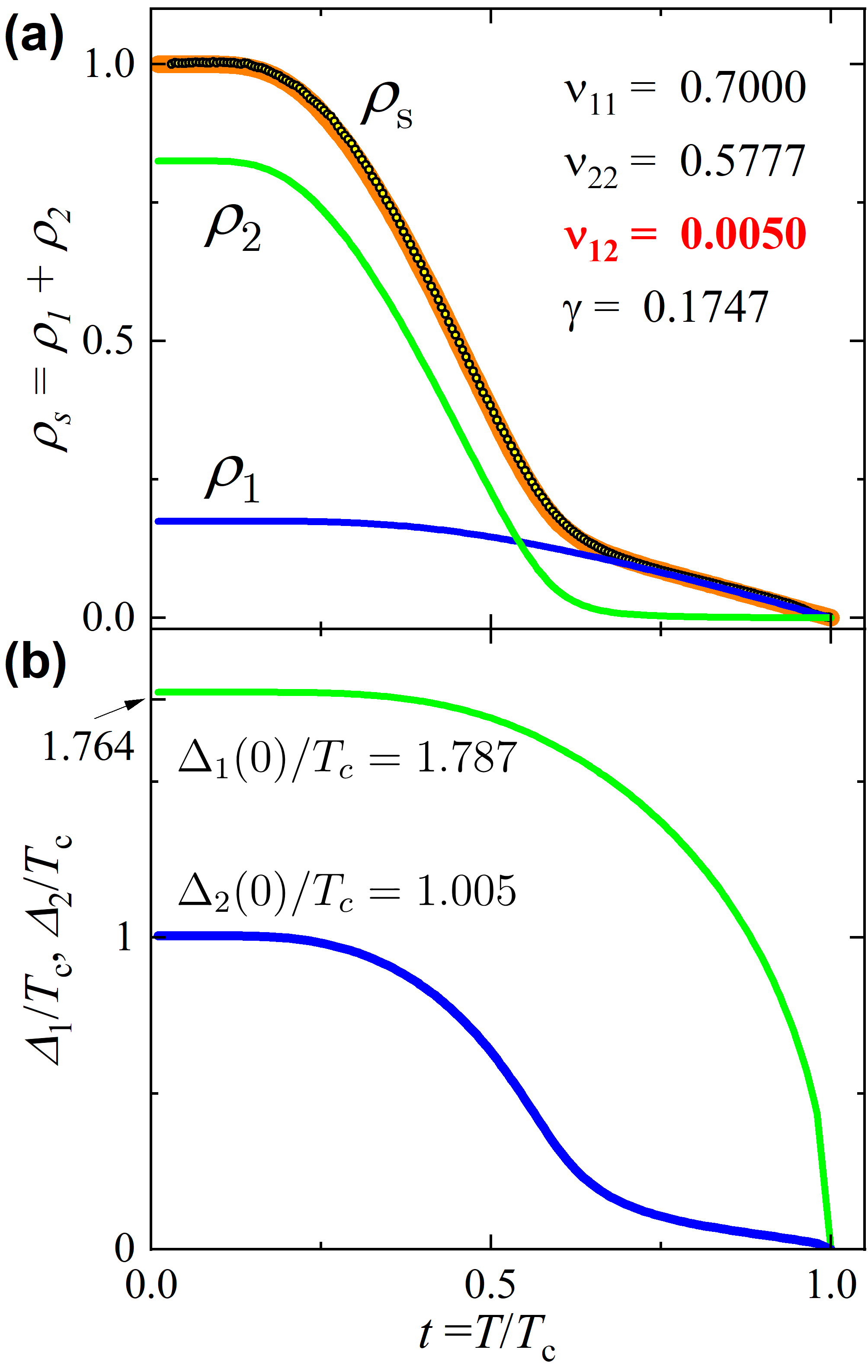}
\caption{(a) Symbols show the superfluid density in pristine sample calculated
from the data shown in Fig.\ref{fig2:lambda}. Blue and green solid
lines show labeled partial superfluid densities, $\rho_{1}$ and $\rho_{2}$,
obtained in the least squares fitting. The thick orange line behind
the data shows an excellent agreement of the data with the fitted
total superfluid density, $\rho_{s}=\gamma\rho_{1}+\left(1-\gamma\right)\rho_{2}$.
Best fit parameters are: $\nu_{11}=0.700$, $\nu_{22}=0.578$, $\nu_{12}=0.005$,
and $\gamma=0.175$. (b) Best fit solutions of the self-consistency
gap equations, Eq.\ref{eq:deltas}. The $T=0$ gap ratios are, $\varDelta_{1}/T_{c}=1.787$
and $\varDelta_{2}/T_{c}=1.005$.}
\label{fig3:SFDfit} 
\end{figure}

Figure~\ref{fig3:SFDfit} (a) shows the normalized superfluid density
of a $\textrm{V}_{3}\textrm{Si}$ crystal in pristine state. A similar
curve for a different crystal, cut from the same master boule, showing
the same two-gap structure, was published in our earlier paper where
a self-consistent $\gamma-$model based on Eilenberger formalism was
introduced \cite{Kogan2009PRB_gamma_model}. In the original $\gamma-$model,
two isotropic $s-$wave gaps are obtained from the solution of the
self-consistency equation, and then all thermodynamic quantities, including
the superfluid density, can be calculated. The model was further generalized
to include anisotropic or even nodal gaps \cite{Li2016}. Here, it
is sufficient to consider the original isotropic approach.

\subsection{The isotropic $\gamma-$model}

The $\gamma-$model considers two bands with Fermi velocities, $v_{i}$,
and the partial densities of states, $n_{i}=N_{i}\left(0\right)/N\left(0\right)$,
where $N\left(0\right)$ is the total density of states at Fermi level,
so that $n_{1}+n_{2}=1$. The dimensionless effective interaction
constants are defined as $\nu_{ik}=N(0)V_{ik}$, where $V_{ik}$ is
the electron-electron interaction matrix. Note that in the original
paper we used $\lambda$ for the interaction matrix. To avoid confusion
with the London penetration depth, here we use $\nu_{ik}$. Also note
that this definition differs from that used in the literature, $g_{ik}=n_{k}\nu_{ik}$.
Our notation has an advantage of being symmetric, $\nu_{ik}=\nu_{ki}$.
Therefore, for two bands, we have three coefficients of the interaction
matrix, two in-band, $\nu_{11}$ and $\nu_{22}$, and an inter-band
coupling, $\nu_{12}$. In the analysis, we perform a least-squares
fit of the experimental superfluid density shown in \ref{fig3:SFDfit}
(a) in Matlab. If all normal-state parameters of a material are known,
$\nu_{ik}$ are the three fitting parameters. They are reduced to
two free parameters by the equation for $T_{c}\left(\nu_{ik}\right)$,

\begin{equation}
1.7638k_{B}T_{c}=2\hbar\omega_{D}\exp(-1/\widetilde{\nu})\label{eq:Tc}
\end{equation}
where we assume conventional electron-phonon mechanism of superconductivity
with $\omega_{D}$ being the Debye frequency. In general, if energy
of bosonic pairing ``glue'' is known, it should be substituted instead
of $\hbar\omega_{D}$ in Eq.\ref{eq:Tc}. The pre-factor comes from
the weak-coupling approximation used in the $\gamma-$model. The effective
interaction constant, $\widetilde{\nu}\left(\nu_{ik}\right)$, is
obtained from the solution of algebraic equations containing all coefficients,
$\nu_{ik}$, see Section II.A of Ref. \cite{Kogan2009PRB_gamma_model}.
To fit the superfluid density, first the self-consistent
gap equation is solved at each temperature. Introducing dimensionless
quantities, $\delta_{i}=(\Delta_{i}/T)/(2\pi t)$, where $t=T/T_{c}$,
the gaps equations are given by

\begin{align}
\delta_{i} & =\sum_{k=1,2}n_{k}\nu_{ik}\delta_{k}\left(\widetilde{\nu}^{-1}-\ln t-A_{k}\right),\nonumber \\
A_{k} & =\sum_{n=0}^{\infty}\left[\left(n+1/2\right)^{-1}-\left(\delta_{k}^{2}+\left(n+1/2\right)^{2}\right)^{-1/2}\right]\label{eq:deltas}
\end{align}
Note that we often set Boltzmann constant, $k_{B}=1$, where it is
obvious, and use it explicitly to emphasize the numerical values or
proper dimensions, e.g., Eq.\ref{eq:Tc}. For a given set of the coupling
constants, $\nu_{ik}$, and partial densities of states, $n_{i}$,
this system can be solved numerically for $\delta_{i}\left(t\right)$
and therefore provide the energy gaps, $\Delta_{i}\left(t\right)=2\pi T\delta_{i}(t)$.
This is a crucial step missing in the so-called $\alpha-$model description
of the two-band superconductivity \cite{Bouquet2001}. While it was
useful early on to explain experimental signatures of two-gap superconductivity
in $\textrm{MgB}_{2}$, the fitting parameters of $\alpha-$model
have little physical meaning. A follow-up study used two-gap functions
pre-calculated from the microscopic theory and showed an excellent
agreement between experimental and theoretical superfluid density,
$\rho_{s}\left(t\right)$ \cite{Fletcher2005}. Indeed, the $s-$wave
$\textrm{MgB}_{2}$ for which all normal-state parameters are known
is a perfect demonstration of the $\gamma-$model where different
quantities are calculated from $\nu_{ik}$ obtained from the fit of
$\rho_{s}\left(t\right)$ \cite{Kogan2009PRB_gamma_model,MgB2Symmetry2019}.

Figure \ref{fig3:SFDfit}(b) shows two gaps calculated self-consistently
from Eq.\ref{eq:deltas}. The individual gap ratios are, $\Delta_{1}/T_{c}=1.787$
and $\Delta_2/T_{c}=1.005$. This should be compared with the results
of microwave surface impedance measurements where similar apparent
two-gap behavior was observed in the superfluid density and the values
of $\Delta_{1}/T_{c}=1.8$ and $\Delta_{2}/T_{c}=0.95$, quite close
to ours, were derived \cite{Nefyodov2005EPL_two-gap_V3Si}. In the absolute
units we obtain, $\Delta_{1}\left(0\right)\approx2.53\;\textrm{meV}$
and $\Delta_{2}\left(0\right)\approx1.42\;\textrm{meV}$. After the
gaps are calculated, the total superfluid density, $\rho_{s}=\gamma\rho_{1}+\left(1-\gamma\right)\rho_{2}$
can be evaluated and fitted to the experimental data. The partial
contributions to the superfluid density are given by \cite{Kogan2009PRB_gamma_model},

\begin{align}
\rho_{i} & =\delta_{i}^{2}\sum_{n=0}^{\infty}\left[\delta_{i}^{2}+(n+1/2)^{2}\right]^{-3/2},\nonumber \\
\gamma & =\frac{n_{1}v_{1}^{2}}{n_{1}v_{1}^{2}+n_{2}v_{2}^{2}}\label{eq:SFD}
\end{align}

\noindent where $v_{i}$ are the Fermi velocities (not to be confused
with Greek $\nu_{i}$ of the interaction matrix). Analyzing Fig.\ref{fig1:BS},
we group bands 1 and 2 into one effective band $I$, and bands 3 and
4 into another band $II$, and we can safely neglect band 5. (Here
we use Roman numerals $I$ and $II$ to index these ``effective''
bands). For the first effective band, we find $\gamma=\left(n_{1}v_{1}^{2}+n_{2}v_{2}^{2}\right)/\sum_{i=1}^{5}n_{i}v_{i}^{2}=0.109$.
If we included band 5, it'd make the difference only in 6$^{th}$
decimal digit. Using $\gamma$ as another fitting parameter, the best
fit of this model to the data gave $\nu_{11}=0.700$ (fixed by $T_{c}$,
Eq.\ref{eq:Tc}), $\nu_{22}=0.578$, $\nu_{12}=0.005$, and $\gamma=0.175$,
with the effective $\widetilde{\nu}=0.350$ (see Eq.\ref{eq:Tc}).
Remarkably, the best-fit value of $\gamma$ is quite close to the
estimate from the electronic band-structure calculations, see Table
\ref{table:BS} where we find $\gamma=0.109$, . This gives confidence
in the model and shows its applicability to describe the superconductivity
in $\textrm{V}_{3}\textrm{Si}$. Naturally, overall, smaller partial
density of states on band $I$, somewhat counter-intuitively, leads
to a larger gap, which is the property of the self-consistent two-band
model \cite{Kogan2009PRB_gamma_model,Kogan2016}. We note that the
possible uncertainty in the experimental value of $\lambda\left(0\right)$
leads to some uncertainty in the fitting parameters, but not large
enough to alter the general conclusion of the relative amplitudes
of the obtained interaction matrix.

\begin{figure}[tb]
\includegraphics[width=0.9\linewidth]{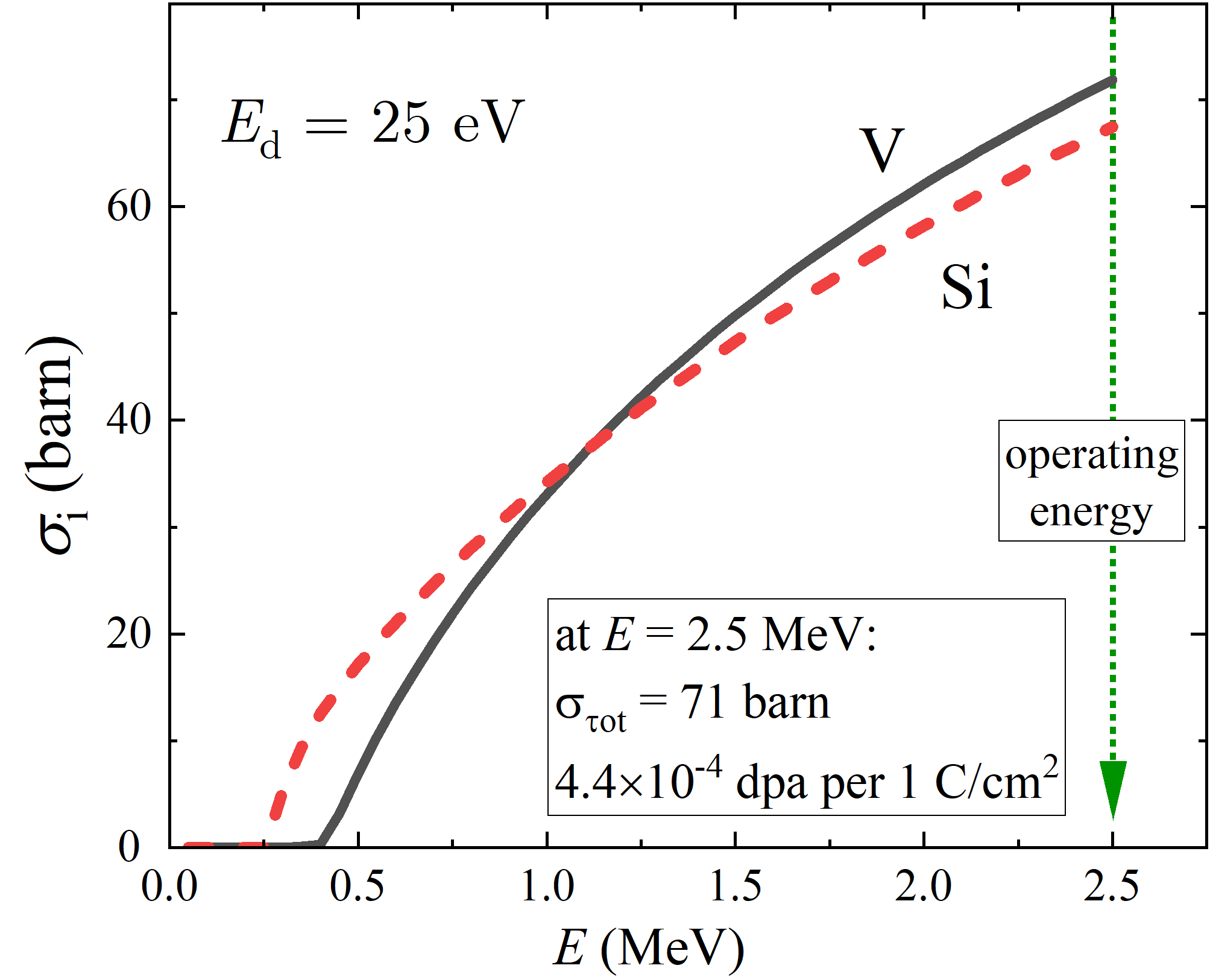}
\caption{Knock-out defects creation cross-sections for vanadium and silicon
ions in $\textrm{V}_{3}\textrm{Si}$ as function of electron energy
assuming the displacement energy threshold, $E_{d}=25\:\mathrm{eV}$.
At the operating energy of 2.5 MeV, the total cross-section is $\sigma=72$
barn, which leads to the estimate of $4.4\times10^{-4}$ displacements-per-atom
(dpa) per 1 C/cm$^{2}$ of the irradiation.}
\label{fig4:SECTE} 
\end{figure}

\subsection{Effect of electron irradiation}

In the last three decades, many studies involving particle irradiation
were performed on various conventional and unconventional superconductors
and there is a vast literature on this topic \cite{Brown1990,Weber1986,Tamegai_2012,Cho2018SST_review_e-irr}
Due to the differences in the rest mass and irradiation temperature,
the number and the morphology of the created defects varies significantly
between different projectile particles. It appears that MeV electrons,
thanks to a small rest mass, transfer just enough energy upon collision
with ions, of the order of tens of eV, to produce well-defined point-like
scattering centers \cite{Ghimire2021}. Much larger energy transfer,
for example from protons, produce many secondary collisions and less
localized damage. A more detailed discussion of electron irradiation
and created defects in solids can be found elsewhere \cite{Damask1963,THOMPSON1969}.

Figure \ref{fig4:SECTE} shows the ion-type-resolved cross-sections
of the defects creation calculated using SECTE (``Sections Efficaces
Calcul Transport d'\'{E}lectrons") software, developed at \'{E}cole Polytechnique
(Palaiseau, France) by members of the ``Laboratoire des Solides Irradi\'{e}s",
specifically for the interpretation of MeV-range electron irradiation
using their Pelletron-type linear accelerator, SIRIUS \cite{SIRIUS}.
Basically, this is a computer-assisted atomic-weights-averaged interpolation
of the ion knock-out cross-sections tabulated by O. S. Oen \cite{Oen1973}.
In the absence of microscopic calculations, we used the commonly assumed
value of the ion displacement energy upon a head-on collision, $E_{d}=25$
eV. The partial cross-sections are very similar, and we expect a roughly
equal number of defects on vanadium and silicon sites. At the operational
energy of 2.5 MeV, the total cross-section is estimated as $\sigma=72$
barn, which means that roughly 1.8 defects are produced per 1000 unit
formulas per 1 C/cm$^{2}$. This is a small number of defects that cannot
change the electronic structure in any appreciable way, and a significant
reduction of $T_{c}$ observed in our experiments must have a different
explanation, such as its pair-breaking nature.

\begin{figure}[tb]
\includegraphics[width=0.9\linewidth]{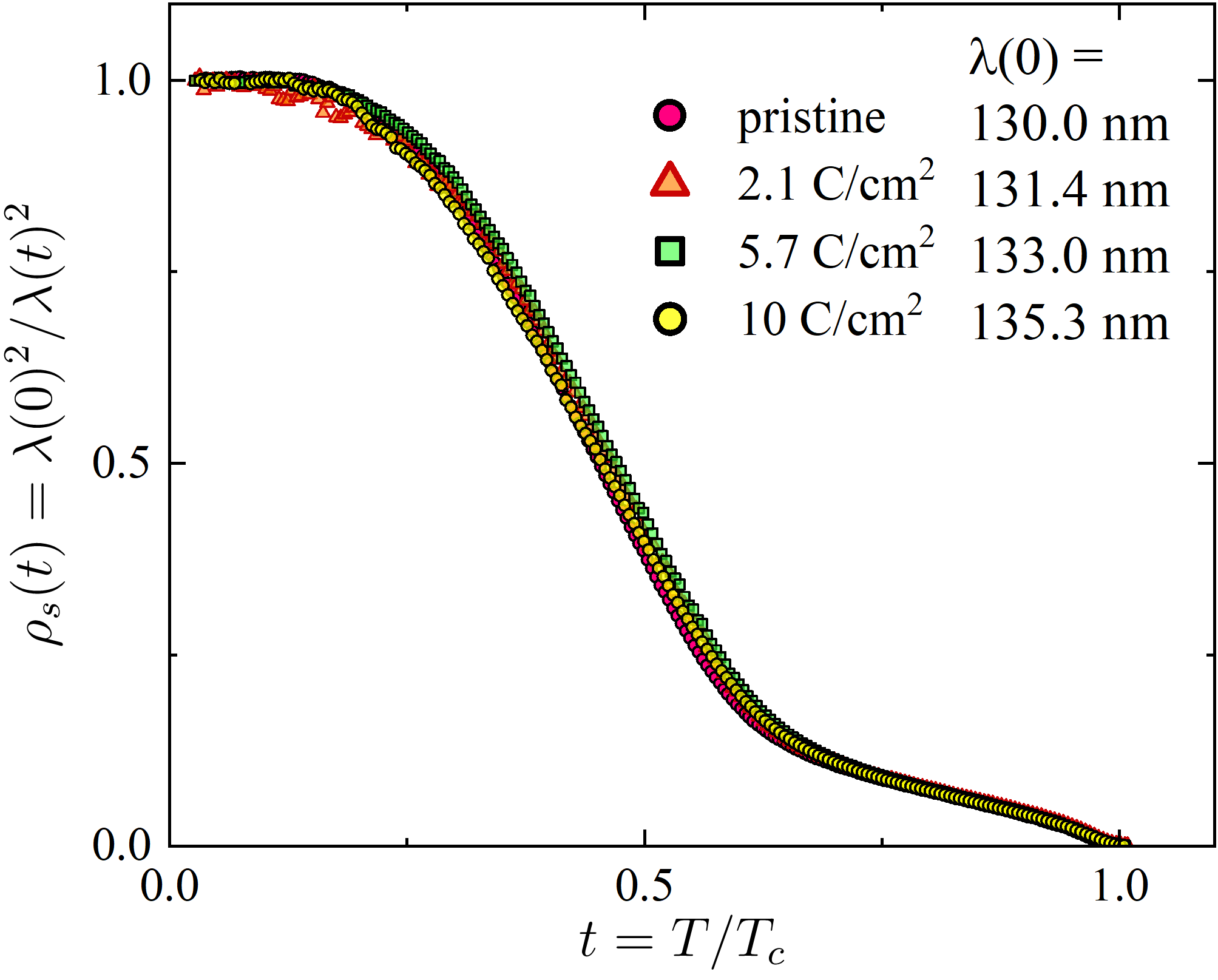}
\caption{Evolution of superfluid density ($\rho_{s}$) upon irradiation. For
the pristine case, we used $\lambda(0)=130$ nm from Ref.~\cite{Wu1994Hyperfine_V3Si_Lambda_0}.
The curves representing irradiated state were calculated with the
penetration depth $\lambda(0)$ values shown in the legend. They were
chosen to collapse the curves on the pristine one.}
\label{fig5:SFDscaled} 
\end{figure}

First, let us examine the effect of electron irradiation on superfluid
density. While we do not know how much $\lambda\left(0\right)$ changes,
we attempted to adjust its value to scale all curves onto a pristine
one. As shown in Fig.\ref{fig5:SFDscaled} this worked rather well
with a small increase of $\lambda\left(0\right)$ values shown in
the legend. This indicates that scattering does not alter the gap
values themselves and, due to very small inter-band coupling, has practically
no effect on the total superfluid density. Each isotropic band follows
the Anderson theorem \cite{Anderson1959a} and the change in $T_{c}$
comes mostly from the inter-band scattering between order parameters
of different magnitude. We note that the relative change of $\lambda\left(0\right)$
can be estimated from Tinkham's widely-used approach \cite{TinkhamBOOK}
that gives for moderate scattering, $\lambda\approx\lambda_{clean}\sqrt{1-\xi_{0}/\ell}$
, where $\xi_{0}\approx60$ nm is the BCS coherence length, and $\ell\approx30$
nm is the electronic mean free path in the pristine state. Both numbers
are estimated for V$_{3}$Si from $T_{c}$, Fermi velocity and resistivity,
see Table \ref{table:BS} and Fig.\ref{fig2:lambda}. As shown in
Fig.\ref{fig2:lambda}, at the maximum irradiation dose, the resistivity
doubles. Therefore, we expect the increase of $\lambda\left(0\right)$
by a factor of about 1.3, which is not large and does not alter our
conclusions, especially considering an apparent scaling shown in Fig.\ref{fig5:SFDscaled}
.

\subsection{Suppression of $T_{c}$ by disorder in a two-band superconductor}

While it is clear that the superfluid density shows a convincing two-distinct-gaps
features implying small inter-band coupling, this still leaves an
unanswered important question of the relative sign of the order parameter
on each band. This is because superfluid density, as well as any other
thermodynamic quantity includes even powers of the gap function,
so that an $s_{\pm}$ state cannot be distinguished from an $s_{++}$
state if the gaps are the same, see, for example, Eq.\ref{eq:SFD}.
The suppression of $T_{c}$ on the other hand is very sensitive to
the overall anisotropy of the order parameter, including a generalized
view when two bands are considered side by side along the common path
on the entire Fermi surface \cite{Kogan2002}. This situation can
be analyzed employing a very useful ansatz that temperature and angular
parts of the order parameter can be separated, $\Delta\left(T,\mathbf{k}_{F}\right)=\Psi\left(T\right)\Omega\left(\mathbf{k}_{F}\right)$,
where $\mathbf{k}_{F}$ is Fermi wave vector and the angular part
obeys the normalization condition for the Fermi surface average, $\left\langle \Omega^{2}\right\rangle _{FS}=1$
\cite{Kogan2002,Anisotropies_PRB2019}. For example, for a single
band $s-$wave, $\Omega=1$ and for a $d-$wave, $\Omega=\sqrt{2}\cos\left(2\varphi\right)$.
For a two-gap superconductor Kogan introduced two gaps each described
by its own angular part $\Omega_{i}$ \cite{Kogan2002}. In this case,
the normalization reads, 
\begin{equation}
\left\langle \Omega^{2}\right\rangle =n_{1}\left\langle \Omega_{1}^{2}\right\rangle +n_{2}\left\langle \Omega_{2}^{2}\right\rangle =1
\label{eq:normalization}
\end{equation}

In the case of an anisotropic gap, even non-magnetic (no spin-flip) scatterers
suppress superconducting transition temperature, $T_{c}$. With spin-flip
scattering both channels reduce $T_{c}$. Openov gives a generalized
Abrikosov-Gor'kov \cite{AbrikosovGorkov1960ZETF} type expression
where gap anisotropy is explicitly taken into account \cite{Openov1997,Openov2004}.
We note that a more general theory of the $T_{c}$ suppression by
disorder scattering, extended to topologically non-trivial superconductors,
is discussed elsewhere \cite{Krenkel2021,Teknowijoyo2018}. Here it
suffice to consider the $\Omega-$ approach, which gives,

\begin{multline}
\ln t_{c}=\psi\left(\frac{g+g_{m}}{2t_{c}}+\frac{1}{2}\right)-\psi\left(\frac{1}{2}\right)\\
-\left<\Omega\right>^{2}\left[\psi\left(\frac{g+g_{m}}{2t_{c}}+\frac{1}{2}\right)-\psi\left(\frac{g_{m}}{t_{c}}+\frac{1}{2}\right)\right]\label{eq:AG}
\end{multline}
where $t_{c}=T_{c}/T_{c0}$ with $T_{c0}$ being the transition temperature
in a pristine state, and $\psi$ is the digamma function. Dimensionless
magnetic and non-magnetic scattering rates are given by

\begin{equation}
g_{(m)}=\frac{\hbar}{2\pi k_{B}T_{c0}}\frac{1}{\tau_{(m)}}\label{eq:scattering-rates}
\end{equation}

\noindent where $\tau$ and $\tau_{m}$ are non-magnetic and magnetic
(spin flip) scattering times, respectively. (Note that original Abrikosov-Gor'kov
theory uses a different definition of the scattering rate, $\rho=\hbar/\left(\pi k_{B}T_{c}\tau\right)$,
with the actual (suppressed) $T_{c}$). The effect of gap anisotropy
can be immediately seen from Eq.\ref{eq:AG} - it contains $\Omega$
in the first power. For a single-band $s-$wave, $\left\langle \Omega\right\rangle =1$
and we obtain $t_{c}=1,$ recovering the Anderson theorem \cite{Anderson1959a}.
For a $d-$wave, $\left\langle \Omega\right\rangle =0$ and we obtain
an expression where both magnetic and non-magnetic impurities suppress
$T_{c}$. It is interesting to note that the critical value for the
complete $T_{c}$ suppression of an $s-$wave order parameter by magnetic
impurities, $g_{m}=0.14$, is exactly half of the value for a $d-$wave
order parameter suppression by non-magnetic impurities, $g=0.28$.

We can now use Eq.\ref{eq:AG} with the two-gaps $\Omega-$approach,
Eq.\ref{eq:normalization} and that $\left\langle \Omega\right\rangle =n_{1}\left\langle \Omega_{1}\right\rangle +n_{2}\left\langle \Omega_{2}\right\rangle $.
Specifically, we consider two isotropic gaps described by constant
values $\Omega_{1}$ and $\Omega_{1}$. In other words, two bands
are represented in a generalized single angular coordinate, - band
1 from $0$ to $2\pi$, and band 2 from $2\pi$ to $4\pi$, each with
its own density of states $N_{i}$, so that the partial densities
of states are $n_{i}=N_{i}/\left(N_{1}+N_{2}\right)$. Introducing
the gap ratio, $r=\Omega_{2}/\Omega_{1}$ and the ratio of the partial
densities of state, $n=n_{2}/n_{1}$=N$_{1}/N_{2}$, we obtain for
the total average in this two-gap model: 
\begin{equation}
\left\langle \Omega\right\rangle ^{2}=\frac{(nr+1)^{2}}{(n+1)\left(nr^{2}+1\right)}\label{eq:OmegaSquared}
\end{equation}
Without magnetic scattering ($g_{m}=0$), the transition temperature
of a two-band superconductor is,

\begin{multline}
\ln t_{c}=\psi\left(\frac{g}{2t_{c}}+\frac{1}{2}\right)-\psi\left(\frac{1}{2}\right)\\
-\frac{(nr+1)^{2}}{(n+1)\left(nr^{2}+1\right)}\left[\psi\left(\frac{g}{2t_{c}}+\frac{1}{2}\right)-\psi\left(\frac{1}{2}\right)\right]\label{eq:AG2band}
\end{multline}

\noindent It is important to emphasize that the superfluid density
as a function of (reduced) temperature and $T_{c}$ are two independent
measurements, which makes the analysis better defined and constrained.

\begin{figure}[tb]
\includegraphics[width=0.9\linewidth]{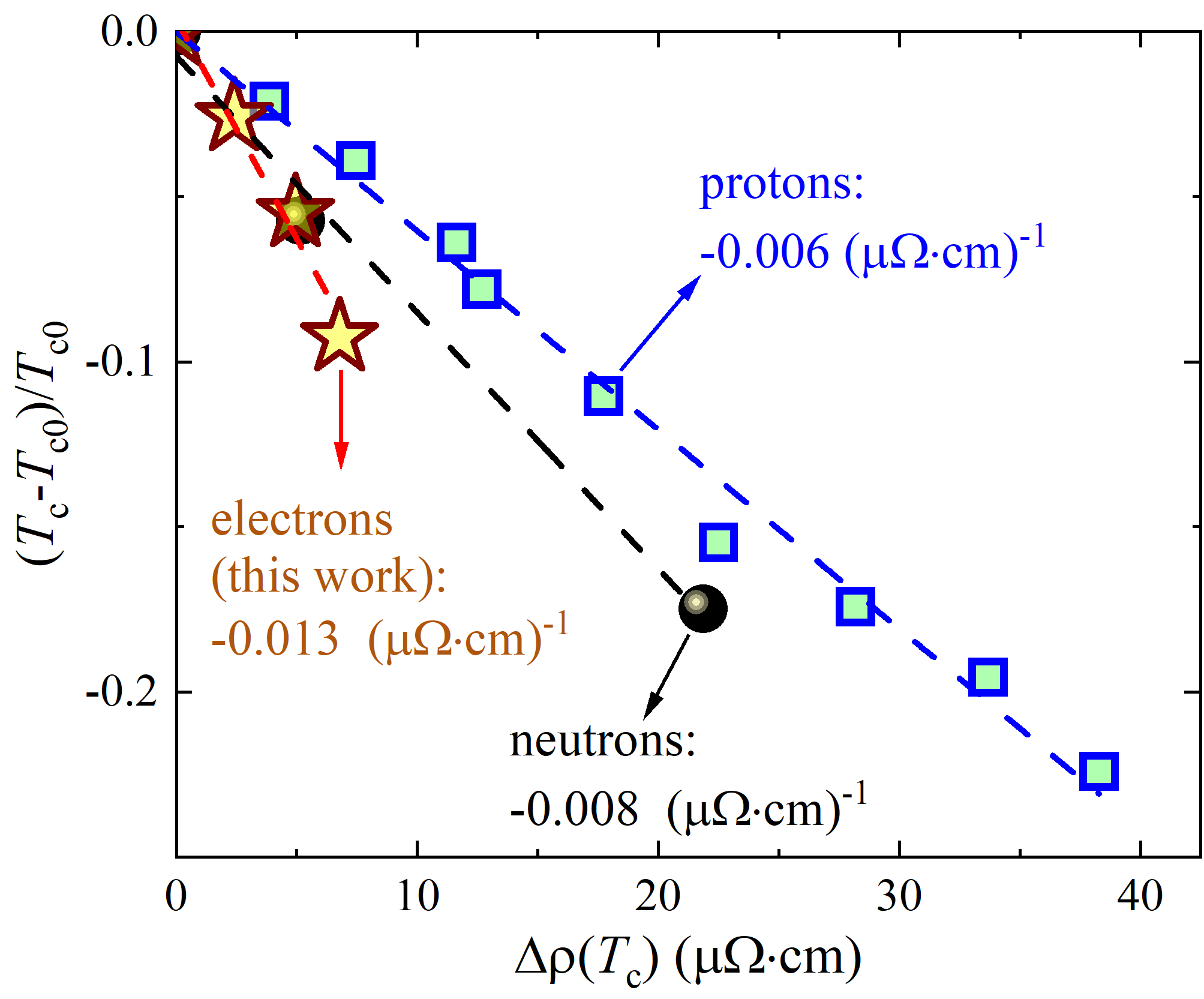}
\caption{Suppression of $T_{c}$ upon different types of particle irradiation.
Normalized $T_{c}$ suppression upon electron irradiation (current
study) is compared with two different previous studies by proton and
neutron irradiations. It is clearly shown that the electron irradiation
is most effective in suppressing $T_{c}$.}
\label{fig6:dDtc/drho} 
\end{figure}

To compare the experimentally observed decrease of $T_{c}$ with our
model, we need a proper parameter characterizing the scattering rate.
The problem is that different sources of disorder produce somewhat
different effects. Figure \ref{fig6:dDtc/drho} compares the relative
change of the transition temperature, $\Delta t_{c}\equiv\left(T_{c}-T_{c0}\right)/T_{c0}$,
in $\text{V}_{3}\text{Si}$ per $1\:\mu\Omega\cdot\textrm{cm}$ of
resistivity increase caused by electron irradiation in the current
study with two previous studies where defects were induced by proton~\cite{Alterovitz1981PRB_proton_V3Si}
and neutron~\cite{Viswanathan1978PRB_neutron_V3Si} irradiation.
The rates of the relative change are, $d\Delta t_{c}/d\rho=-0.013$
$\left(\mu\Omega\cdot\textrm{cm}\right)^{-1}$ (electron irradiation),
-0.008 $\left(\mu\Omega\cdot\textrm{cm}\right)^{-1}$ (neutron), and
-0.006 $\left(\mu\Omega\cdot\textrm{cm}\right)^{-1}$ (proton). Due
to their small rest mass and matching range of the energy transfer
(1-100 eV), electrons produce the most efficient point-like defects and
have the largest suppression rate. A similar trend is observed in other
materials, for example, well-studied iron-based superconductors \cite{Cho2018SST_review_e-irr,Ghimire2021}.
On the other hand, the observed rates are not too different, roughly
0.01 $\left(\mu\Omega\cdot\textrm{cm}\right)^{-1}$, and we can put
it in a perspective by comparing with other superconductors. For that,
we need to calculate the dimensionless scattering rate, Eq.\ref{eq:scattering-rates}.
In our case of measured $\lambda\left(T\right)$ and $\rho\left(T\right)$
the simplest estimate of the scattering time, $\tau$, is via the
London and Drude electrodynamics, $\tau\left(T_{c}\right)=\mu_{0}\lambda_{clean}^{2}\left(0\right)/\rho\left(T_{c}\right)$.
Note that clean-limit value, $\lambda_{clean}\left(0\right)$, needed
for the density of states in the normal metal, enters this estimate,
whereas (normal metal) scattering time comes from resistivity.
This approach is well justified in isotropic $s-$wave superconductors
and $s_{++}$ compounds assuming that the gap smearing caused by the
modest amounts of non-magnetic disorder is much smaller than the gap
amplitudes.

In our case, we can use Tinkham's widely-used approach \cite{TinkhamBOOK}
that gives for moderate scattering, $\lambda\approx\lambda_{clean}\sqrt{1-\xi_{0}/\ell}$
, where $\xi_{0}\approx60$ nm is the BCS coherence length, and $\ell\approx30$
nm is the electronic mean free path in the pristine state. Both numbers
are estimated for V$_{3}$Si from $T_{c}$, Fermi velocity and resistivity,
see Table \ref{table:BS} and Fig.\ref{fig2:lambda}. As shown in
Fig.\ref{fig2:lambda}, at the maximum irradiation dose, the resistivity
doubles. Therefore, we expect the increase of $\lambda\left(0\right)$
by a factor of about 1.3. This is an insignificant change to alter the main features reported here - the
exponential attenuation at low temperatures and a higher temperature
kink signaling of two barely-coupled gaps of different magnitude.
This is further confirmed by the apparent scaling of the superfluid
density for all doses of electron irradiation, Fig.\ref{fig5:SFDscaled}.

The experimental dimensionless scattering rate can be estimated as,

\begin{equation}
g\approx\frac{\hbar}{2\pi k_{B}\mu_{0}}\frac{\rho(T_{c})}{T_{c0}\lambda_{clean}(0)^{2}}\label{eq:gExp}
\end{equation}

Note that we measure resistivity change with respect
to the pristine sample to subtract inelastic scattering, but this also removes background impurity scattering in samples before irradiation. Fortunately, judging by very low pinning, this correction is negligible
\cite{Kogan_1997}. Also, note that this dimensionless rate contains unmodified $T_{c0}$, which is different from the original Abrikosov-Gor'kov definition \cite{AbrikosovGorkov1960ZETF}.

\begin{figure}[tb]
\includegraphics[width=0.9\linewidth]{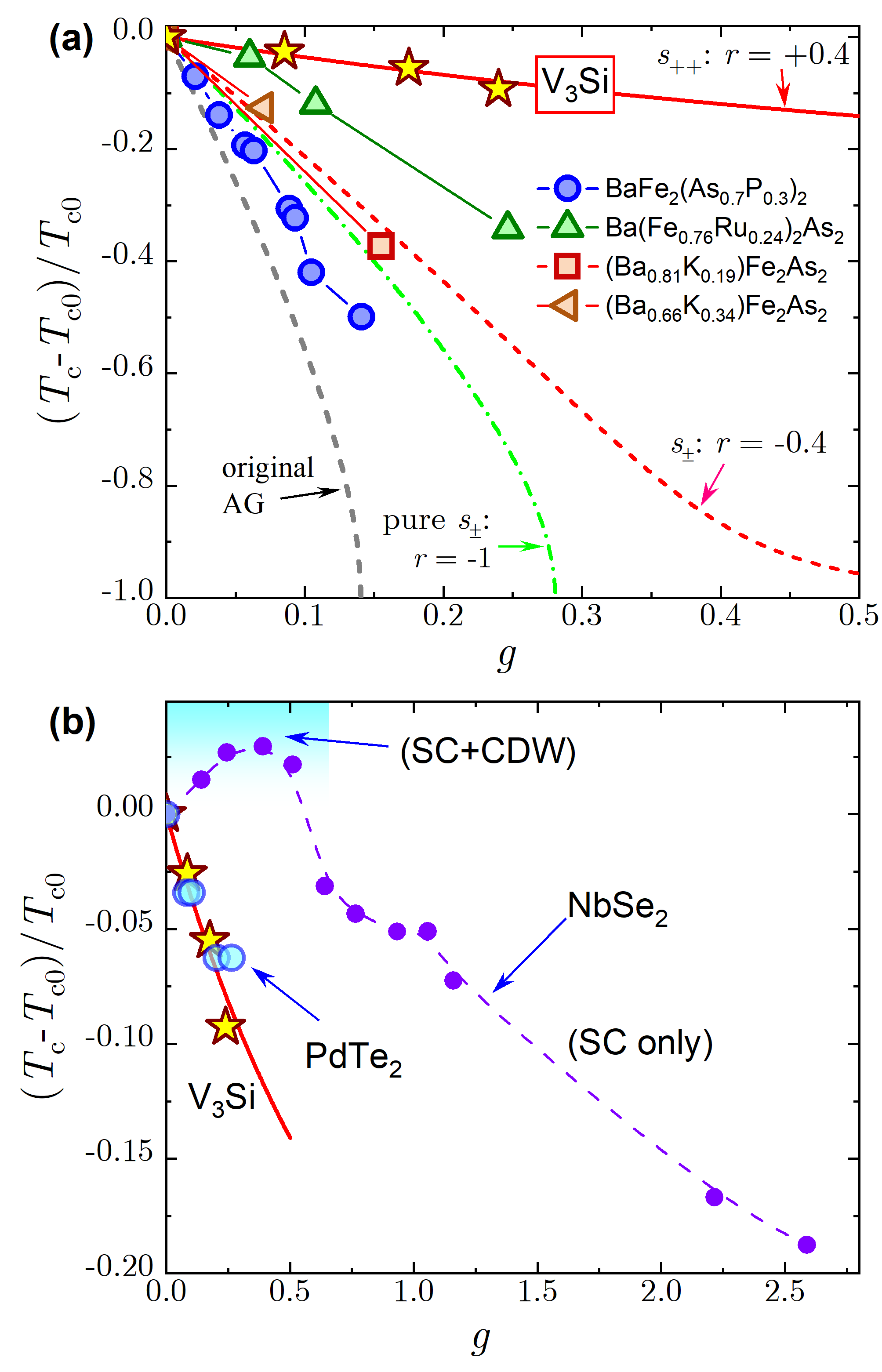} \caption{Normalized change of superconducting transition temperature, $\Delta t_{c}\equiv\left(T_{c}-T_{c0}\right)/T_{c0}$
as function of the dimensionless scattering rate, $g$. (a) comparison
of $\text{V}_{3}\text{Si}$ single crystal with known nodeless and
nodal $s_{\pm}$ superconductors shown in the legend. The theoretical
curves from Eq.\ref{eq:AG2band} are shown by lines. Clearly, all
sign-changing $s_{\pm}$ superconductors show suppression rate larger
than in $\text{V}_{3}\text{Si}$. (b) similar comparison with another
$s_{++}$ two-band superconductor, $\text{NbSe}_{2}$ \cite{Cho2018}
and unconventional Dirac semi-metal compound, $\text{PdTe}_{2}$ \cite{Teknowijoyo2018,Timmons2020}.
As soon as CDW is suppressed, $\text{NbSe}_{2}$ shows a similar suppression
rate as $\text{V}_{3}\text{Si}$.}
\label{fig7:DTc-vs-g} 
\end{figure}

Figure \ref{fig7:DTc-vs-g} shows the normalized change of superconducting
transition temperature, $\Delta t_{c}\equiv\left(T_{c}-T_{c0}\right)/T_{c0}$,
as function of the dimensionless scattering rate, $g$. Figure \ref{fig7:DTc-vs-g}(a)
compares $\text{V}_{3}\text{Si}$ single crystal with known nodeless
and nodal $s_{\pm}$ superconductors, isovalently substituted $\text{Ba}(\text{Fe}_{0.76}\text{Ru}_{0.24})_{2}\text{As}_{2}$
and $\text{BaFe}_{2}(\text{As}_{0.7}\text{P}_{0.3})_{2}$, and hole-
doped underdoped $\text{Ba}_{0.81}\text{K}_{0.19}\text{Fe}_{2}\text{As}_{2}$
and optimally-doped $\text{Ba}_{0.66}\text{K}_{0.34}\text{Fe}_{2}\text{As}_{2}$
(BaK122) \cite{Cho2018SST_review_e-irr}. The theoretical curves from
Eq.\ref{eq:AG2band} are shown by different lines. Clearly, all sign-changing
$s_{\pm}$ superconductors show suppression rate significantly higher
than in two-gap $\text{V}_{3}\text{Si}$, which is consistent with
$s_{++}$ theoretical curve for gap ratio, $r=+0.4$, while the same
gap ratio, but of opposite signs, $r=-0.4$, is close to BaK122 data.
Interestingly, a nodal multi-band $s_{\pm}$ superconductor, $\text{BaFe}_{2}(\text{As}_{0.7}\text{P}_{0.3})_{2}$,
shows even greater rate of $T_{c}$ suppression, most likely because
in this case the inter-band and in-band scattering channels are both
pair-breaking. Figure \ref{fig7:DTc-vs-g}(b) compares $\text{V}_{3}\text{Si}$
with another $s_{++}$ two-band superconductor, $\text{NbSe}_{2}$
\cite{Cho2018}, and unconventional Dirac semi-metal compound, $\text{PdTe}_{2}$
\cite{Teknowijoyo2018,Timmons2020}. In $\text{NbSe}_{2}$, the situation
is complicated by the charge-density wave (CDW), whose competition
with superconductivity (SC) leads to the initial increase of $T_{c}$.
However, as soon as CDW is destroyed by disorder, further suppression
of $T_{c}$ is quite similar to our subject compound, $\text{V}_{3}\text{Si}$
\cite{Cho2018}. The second compound, unconventional $\text{PdTe}_{2}$
shows the rate of suppression quite similar to $\text{V}_{3}\text{Si}$.
Moreover, it also has a fully-gapped Fermi surface leading to exponential
attenuation of the penetration depth. However, peculiarities of the
electronic band structure of $\text{PdTe}_{2}$ support unconventional
pairing mechanism \cite{Teknowijoyo2018,Timmons2020}, whereas $\text{V}_{3}\text{Si}$
does not have such topological features and is consistent with BCS-type
two-gap superconductivity. This is a good example showing that measurements
alone cannot answer objectively. They must be supported by
theoretical analysis.

\section{Conclusions}

We used controlled point-like disorder induced by 2.5 MeV electron
irradiation at different doses to study superconducting order parameter
in a $\text{V}_{3}\text{Si}$ single crystal. Simultaneous measurements
of London penetration depth and superconducting transition temperature,
$T_{c}$, set stringent experimental boundaries on possible superconducting
states. Specifically, we observe: (1) exponentially attenuated low-temperature behavior of $\lambda\left(T\right)$ (which means a fully gapped Fermi surface); (2) a kink at higher reduced temperatures (signaling of two barely-coupled gaps); (3) a significant shift of $T_{c}$ (gaps of different amplitude). The discussed analysis is applicable for
any choice of $\lambda\left(0\right)$. Using a two-band analysis
for both quantities, $\rho_{s}\left(T\right)$ and $\Delta T_{c}$,
we conclude that $s_{++}$ pairing with two barely-coupled gaps of
different amplitudes, $\Delta_{1}\left(0\right)\approx2.53\;\textrm{meV}$
and $\Delta_{1}\left(0\right)\approx1.42\;\textrm{meV}$, provide
an excellent fit and overall self-consistent description of the experiment.
This makes $\text{V}_{3}\text{Si}$ the earliest (superconductivity
discovered in 1953) proven $s_{++}$ superconductor, preceding $\text{MgB}_{2}$
(superconductivity discovered in 2001) by half a century.
\begin{acknowledgments}
We thank David Christen for providing excellent single crystals well-characterized in his earlier papers.  This work was supported by the
US Department of Energy (DOE), Office of Science, Basic Energy Sciences,
Materials Science and Engineering Division. Ames Laboratory is operated
for the US DOE by Iowa State University under contract DE-AC02-07CH11358.
The authors acknowledge support from the EMIR\&A French network (FR
CNRS 3618) on the ``SIRIUS" platform under proposal \# 18-5155.
We thank the whole SIRIUS team, O. Cavani, B. Boizot, V. Metayer, and J.
Losco, for operating electron irradiation facility. 
\end{acknowledgments}

%

\end{document}